\documentstyle[twocolumn,aps,psfig]{revtex}

\begin{document}
\tolerance 50000

\draft

\twocolumn[\hsize\textwidth\columnwidth\hsize\csname @twocolumnfalse\endcsname

\title{The Magnetization of $Cu_2(C_5H_{12}N_2)_2Cl_4$ : A Heisenberg
Spin Ladder System.}

\author{C.A. Hayward$^{1}$, D. Poilblanc$^{1}$ and L.P. L\'evy$^{2}$
}
\address{
$^{1}$Lab. de Physique Quantique, Universit\'e Paul Sabatier, 
31062 Toulouse, France.\\ 
$^{2}$High Magnetic Field Lab. CNRS/MPI-FKF, BP 166, F-38042 Grenoble Cedex 9,
France.\\
}

\date{April 96}
\maketitle

\begin{abstract}
\begin{center}
\parbox{14cm}{
We study the magnetization of a Heisenberg spin ladder using exact 
diagonalization techniques, finding three distinct magnetic phases.
We consider the results in relation to the experimental behaviour 
of the new copper compound 
$Cu_2(C_5H_{12}N_2)_2Cl_4$  and deduce that the compound is well
described by such a model with a ratio of `chain' to `rung' 
bond strengths ($J/J^\prime$)
of the order of $0.2$, consistent with results from the magnetic 
susceptibility.
The effects of temperature, spin impurities and additional diagonal bonds 
are presented and we give evidence that these diagonal
bonds are indeed of a ferromagnetic nature.  
}
\end{center}
\end{abstract}

\pacs{
\hspace{1.9cm}
}
\vskip2pc]

There has been considerable recent theoretical interest in coupled chain 
systems for a variety of reasons: Firstly the systems provide an interesting
step from the relatively well understood one-dimensional behaviour towards 
two-dimensional systems (i.e. a dimensional crossover); A second reason
for interest lies in the unusual and exotic behaviour exhibited by spin ladder 
systems, for example a spin gap \cite{heisladder}
and on doping, hole pairing
and a finite superfluid density \cite{HPNSH};
A third, and indeed the dominant, motivation for this article lies in the 
increasing number of compounds which can be well described by considering the
behaviour of strongly correlated electrons confined to coupled chains. The 
compounds $(VO_2)P_2O_7$ \cite{compounds1} and $SrCu_2O_3$ \cite{compounds2}
may be described by ladder spin systems and recently doping has been 
achieved in $La_{1-x}Sr_xCuO_{2.5}$ \cite{compounds3}. In this article 
we shall concentrate on the magnetic behaviour of the new copper compound
$Cu_2(C_5H_{12}N_2)_2Cl_4$ \cite{compound} which in contrast to the other 
examples, exhibits 
a spin gap which is relatively small, thereby
allowing a study of the magnetic effects with a relatively modest magnetic
field. We will show that
the magnetization of the material is well described by the Heisenberg 
model on a ladder system.

\begin{figure}
\psfig{figure=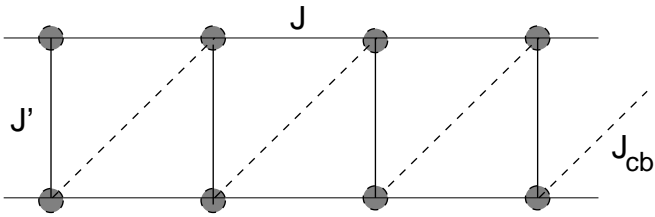,width=\columnwidth,angle=-90}
\caption{
The $Cu$-$Cu$ super-exchange paths in the compound
$Cu_2(C_5H_{12}N_2)_2Cl_4$ (taken from
Chaboussant {\it et al}).
\label{f1}
}
\end{figure}

Recent experimental work on 
$Cu_2(C_5H_{12}N_2)_2Cl_4$,
including magnetization, susceptibility 
and spin resonance experiments, 
has been presented by Chaboussant {\it et al}
\cite{compound}. The material is thought to consist of effectively 
isolated coupled chains as shown in figure \ref{f1}.
Superexchange gives rise to a coupling 
along the chains (strength $J$) and an interchain coupling 
(strength $J^\prime $); there is an additional 
diagonal coupling $J_{cb}$. We shall firstly neglect $J_{cb}$ since its
relative strength is  believed to
be small although 
we shall include it later in the paper. 
Using the susceptibility data,  perturbation theory 
and a high temperature series expansion,
Chaboussant {\it et al} have deduced a bond ratio
$J^\prime /J \sim 5.5$. 

The Hamiltonian we shall use to describe the compound is the Heisenberg 
model on a ladder ($2\times L$) system, defined by 
\begin{eqnarray}
   {\cal H}=
   J^\prime \; \sum_{j} 
   {\bf S}_{j,1} \cdot {\bf S}_{j,2} 
   +J\;\sum_{\beta,j} 
   {\bf S}_{j,\beta} \cdot {\bf S}_{j+1,\beta}\cr 
   + \sum_{\beta,j,\alpha,\gamma} g_{\alpha\gamma} \mu_B
   H^\alpha  S^\gamma_{j,\beta}
\label{hamiltonian} 
\end{eqnarray}
\noindent
where  $\beta$ (=1,2)
labels the two legs of the ladder (oriented along the 
$x$-axis), $j$ is a rung index ($j$=1,...,$L$) and $J$ and $J^\prime$ are
the bond strengths along and between the ladders respectively. The
final term represents an applied field in the direction $\alpha$; we
simplify this term to $g\mu_B\sum H_zS_z$ 
although we should  note that anisotropic
effects may have minor but observable effects. 

The behaviour of this Hamiltonian  in zero 
applied field ($H=0$) and at zero temperature
is now relatively well understood 
\cite{heisladder} and is perhaps best
understood by first considering the limit $J=0$. 
In this case, the ground state  has total spin zero and 
is formed by creating a singlet bond 
on each rung; excitations require 
one of these singlet bonds to be broken to form a triplet at 
an energy cost $J^\prime$. 
This gapped state persists with the introduction of interchain
coupling $J$ and  the
triplets can propogate and form a coherent band with dispersion 
$J^\prime +J \cos{k}$.   
A gap, and the associated dispersion, has been observed experimentally
in the compound we are considering \cite{compound} and it is
believed that in zero applied field
the energy spectrum of the system consists of a total-spin-zero ground
state, a gap to the first excited state (triplet) and at higher
energies, (bands of) states with even larger values of total spin. 

In this paper then, we will consider the effects of an applied field on
the system i.e. considering the magnetization curve. In the first section, we
calculate ${\cal M}(H)$ at zero temperature, and by comparison with experiment
we deduce the ratio $J/J^\prime$ relevant to the compound. We then
consider the effects of a finite temperature, deduce the relevance
of random spin impurities, and calculate the effect of introducing
the small diagonal interaction.
Finally, we discuss other possible relevant
factors. 

The technique we have used in this study is Lanczos exact diagonalization 
on $2\times 12$ and $2\times 16$ ladder geometries with periodic 
boundary conditions. We have  considered the momentum 
of  the states ($k_x={2\pi\over L}m$ where $m$ is an integer) and also 
the parity of the states 
under a reflection in the symmetry axis along the ladder  
(even ($R_x=1$) or odd ($R_x=-1$)). 
Since the hamiltonian commutes with the component of total spin in the 
$z$-direction ($S_z$), we may consider the subsets
of $S_{z}$ individually. For a specific value of applied field, 
we consider the lowest energy state for each subset and can then 
easily apply the field dependent term of the hamiltonian $\propto  -H.S_{z}$.
With increasing applied field, states with larger $S_{z}$ become 
more favourable  and  
we obtain a `staircase'
of states in the magnetization curve
until saturation. 

\begin{figure}
\psfig{figure=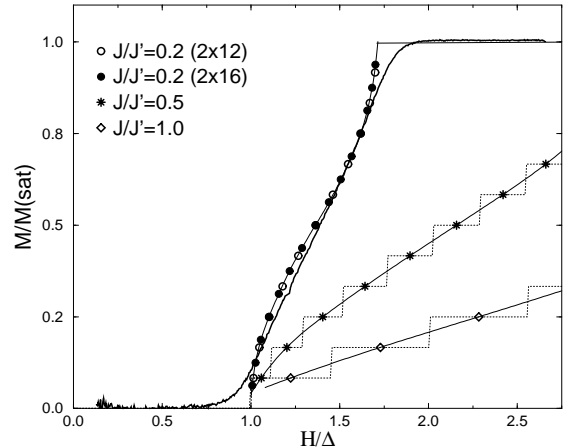,width=\columnwidth,angle=-90}
\caption{
Magnetization as a function of applied field.
${\cal M}(sat)$ is the saturated
value of the magnetization and $\Delta$ is the singlet-triplet gap  
with zero applied field.
Results are shown for $J/J^\prime$ equal to 0.2, 0.5 and 1.0. 
along with the experimental data (bold line).
\label{f2}
}
\end{figure}

In figure \ref{f2} we show results from the $2\times 12$ system for
various values of the ratio $J/J^\prime$
(and also 
the case $J/J^\prime =0.2$
as an example of the $2\times 16$ system).
In addition, we plot the experimental results for the lowest available 
temperature (0.42K). The magnetization is normalized in such a way that
the saturated system has magnetization unity and the applied field is 
normalized such that the value at which magnetization
becomes non-zero is unity. 
For the cases with  $J/J^\prime $ equal  to 
$0.5$ and $1.0$,  we have plotted the staircase structure 
resulting from the finite system; using the midpoints of the `steps'
we have drawn  a smooth magnetization curve. 
For the case $J/J^\prime =0.2$, 
the results from the two system sizes are almost identical indicating 
that the finite size effects are very small (this is also true for 
other $J/J^\prime$).

Immediately we notice that two critical fields may be defined:
For an applied field $H<H_{c1}$ the magnetization remains zero (in
a singlet ground state); the field $H_{c1}$ corresponds to the 
singlet-triplet gap at zero applied field; Then, with increasing magnetic 
field $H_{c1}<H<H_{c2}$, the magnetization increases until it reaches its 
saturation value at $H_{c2}$. 
At this point it is worthwhile mentioning the work of Affleck \cite{IA}
concerning gapped, integer spin antiferromagnetic chains: In an axially
symmetric situation the ground state above $H_{c1}$ 
may be considered as a condensate of low energy bosons; varying the field
varies the chemical potential of the bosons and the boson number corresponds
to the magnetization. In the limit of zero boson density ($H\mapsto H_{c1}$
from above)
the magnetization is shown to behave as
${\cal M}(H)\propto \sqrt{H-H_{c1}}$ and
there is a power law decay in the 
staggered magnetization orthogonal to the applied field \cite{note1}:
For $H_{c1}<H<H_{c2}$ the spins exhibit a 
canted spin structure with a uniform magnetic moment in the direction
of the applied field. The square root singularity appears consistent 
with the theoretical behaviour in figure \ref{f2}. 

We emphasise that the theoretical results with $J/J^\prime =0.2$
are extremely close to the experimental data, and this ratio is consistent 
with that deduced by Chaboussant {\it et al} by analysing susceptibility data.
Notice however  the rounding of the experimental data 
in the region of the critical fields $H_{c1}$ and $H_{c2}$ and the stronger
singular behaviour in the theoretical results. The aim of the 
remainder of this article is to discuss  the origin of these effects 
and we
firstly extend our results to take into account the finite temperature,
specifically looking at the rounding behaviour close to $H_{c2}$. 

It is easy to calculate the complete spectrum of energy levels (by
considering each subset of $S_{z}$ separately) and 
consequently the thermodynamic 
quantities can be calculated. For a specific value of the applied 
field the magnetization is defined by 
\begin{eqnarray}
   {\cal M}=
   {\sum_{n,S_z}\exp{(-\beta E_n^{S_z})}S_z\over
   \sum_{n,S_z}\exp{(-\beta E_n^{S_z})}}
\label{magtemp} 
\end{eqnarray}
\noindent
where  
$E_n^{S_z}$ is the energy of the nth eigenvalue of the hamiltonian 
(equation \ref{hamiltonian}) with a
$z$-component of spin $S_z$. Since we are only interested in the region 
of the magnetization curve close to $H_{c2}$, we restrict the
summation over $S_z$ to $S_z\ge 8$ (for the $2\times 12$ ladder); states with
smaller $S_z$ contribute only minimally in this region (this has been checked) and 
also subsets with $S_z<8$ include many more states and computational
limitations become important. 

\begin{figure}
\psfig{figure=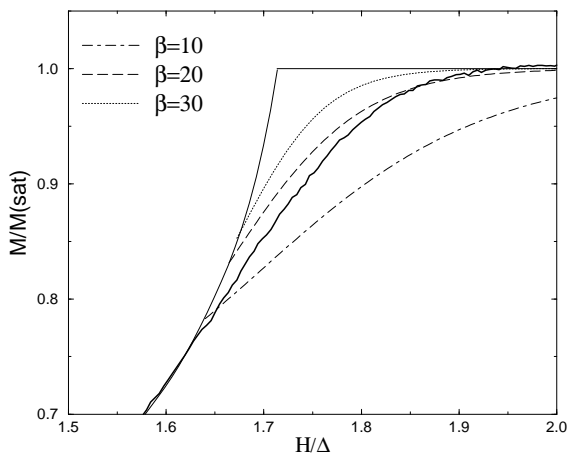,width=\columnwidth,angle=-90}
\caption{
Magnetization as a function of applied field for various values of 
temperature ($1/ \beta$). 
The experimental data (0.42K $\mapsto \beta\sim 30$) is shown as a bold line. 
\label{f3}
}
\end{figure}

The resulting magnetization curve for various values of $\beta$ 
is shown in figure \ref{f3}; we concentrate on $J/J^\prime =0.2$ 
since this corresponds closely to the effective ratio in 
the compound. 
It is immediately obvious that the effect of temperature is to cause a 
rounding of the magnetization curve, much as observed experimentally. 
From experimental considerations,
we would expect a temperature
corresponding to
$\beta\sim {J^\prime \over T} \sim {13.2\over 0.42} \sim 31$  
(where the 13.2 originates from the fact that we have
normalised such that $J^\prime$ is unity, see reference 
\cite{compound}). Therefore, whilst temperature
does indeed effect the magnetization curve in the vicinity of the critical
fields, there must be some other factor to explain the small  discrepancy
between the theory and experiment.

The next step is the inclusion of random spin
impurities, i.e. including a term in the hamiltonian to describe 
the interaction of the local spins with random magnetic fields. 
The additional term included in the hamiltonian has the form
\begin{eqnarray}
   {\cal H}_{imp}=
   \sum_{j,\beta}{w_{j,\beta}S_{j,\beta}^z}
\label{impurity} 
\end{eqnarray}
\noindent
where $w_{j,\beta}$ is the impurity strength at the site $j,\beta$ (as 
defined in the initial hamiltonian), chosen randomly between 
$-w/2$ and $w/2$ and $S_{j,\beta}^z$ is the $z$ component of spin 
on that site.
In order to conserve  the reflection symmetry, we choose
the impurity  weight of the two sites on a particular rung to be equal 
($w_{j,1}=w_{j,2}$). 
The inclusion of random spin impurities breaks the
translational symmetry of the system ($k_x$ is no longer a good quantum number)
and we also note that it is necessary  to average over several
realizations of the 
disorder due to statistical fluctuations.

\begin{figure}
\psfig{figure=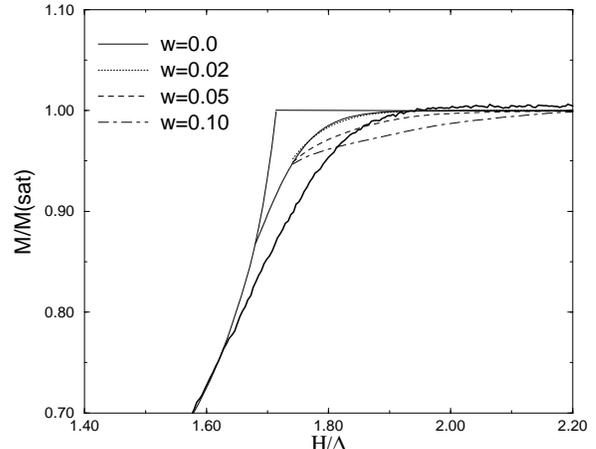,width=\columnwidth,angle=-90}
\caption{
Magnetization as a function of applied field 
including random spin impurities. 
Various impurity weights are shown 
for a temperature corresponding to  
$\beta=30$.
The experimental data is shown as a bold line.
\label{f4}
}
\end{figure}

In Figure \ref{f4} we show the effects of including  random spin 
impurities with various weights ($w$). The dominant effect of the impurities
appears to be an increase in the critical field $H_{c2}$; The effect on
the rounding of the 
magnetization curve does not however appear to be the origin of the small
discrepancy in shape of ${\cal M}(H)$
between the  experimental and theoretical results. 

\begin{figure}
\psfig{figure=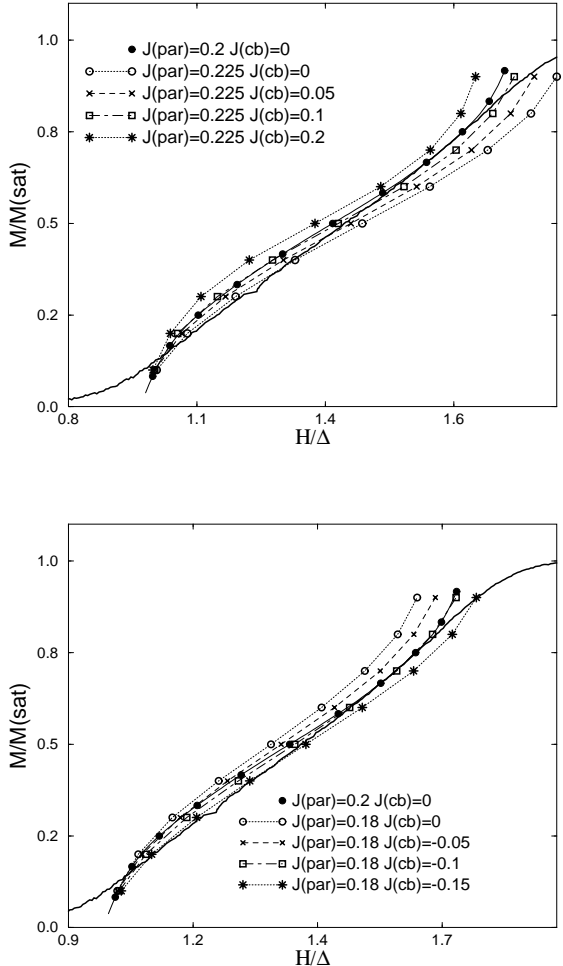,width=\columnwidth,angle=-90}
\caption{
Magnetization as a function of applied field 
including diagonal cross bonds $J_{cb}$ of various strengths.  
Figure 5a corresponds to $J=0.225$ and positive $J_{cb}$ whilst 
figure 5b corresponds to $J=0.18$ and negative $J_{cb}$.  
The experimental data is shown as a bold line and the result 
corresponding to 
$J=0.2$ $J_{cb}=0$ is also included for comparison.
\label{f5}
}
\end{figure}

As a final investigation, we examine the possible importance of a diagonal 
cross bond ($J_{cb}$) as shown in figure \ref{f1}.  
To include this affect we add the following term to the 
hamiltonian 
\begin{eqnarray}
   {\cal H}_{cb}=
   J_{cb} \; \sum_{j} 
   {\bf S}_{j,1} \cdot {\bf S}_{j+1,2} 
\label{hcb} 
\end{eqnarray}
\noindent
As suggested experimentally \cite{compound}, we  take $J_{cb}$ to be
smaller  than 
both $J$ and $J^\prime$.
In figures \ref{f5}a and b we show the magnetization curve for various sets
of parameters, the choice of which has been led by the desire
to keep $H_{c2}/H_{c1}$ close to the experimentally deduced value
(increasing $J_{cb}$ increases the effective interchain coupling and
to keep this ratio constant, $J$ must simultaneously be increased).
The results were calculated for zero temperature
on a $2\times 10$ system (it has
been verified that finite size effects are negligible). As 
previously, 
the results are normalised such that $J^\prime$ is unity, saturated 
magnetization has value unity, and $H_{c1}$ is unity. 
In figure \ref{f5}a we show the experimental data and the theoretical
results corresponding to $J=0.2$ ($J_{cb}=0$) as shown previously,
and $J$=0.225 with the cross bond strength ranging from 0 to 0.2.
A similar plot  (figure \ref{f5}b) shows data corresponding to
$J=0.18$ and {\it negative} cross bond strength ranging from 0 to -0.15.

Several interesting features are apparent in the results.
Firstly, introducing the cross bond interactions does not  
affect the overall behaviour of the magnetization curve. 
The major effects are to firstly change $H_{c1}$
(this effect is not apparent due to the normalization), and secondly to
change the shape of the magnetization curve slightly.
Positive $J_{cb}$ infact changes the shape 
away from the experimental behaviour, seeming to 
increase the singularity behaviour. Surprisingly, a negative 
$J_{cb}$ appears to shift the theoretical curve closer to the experimental
behaviour, with say $J=0.18$ $J_{cb}=-0.1$ being a reasonable parameter choice.
The results seem therefore to suggest a ferromagnetic diagonal interaction,
and further investigations of the orbital behaviour in the compound are 
required to deduce if this is reasonable. 
\begin{figure}
\psfig{figure=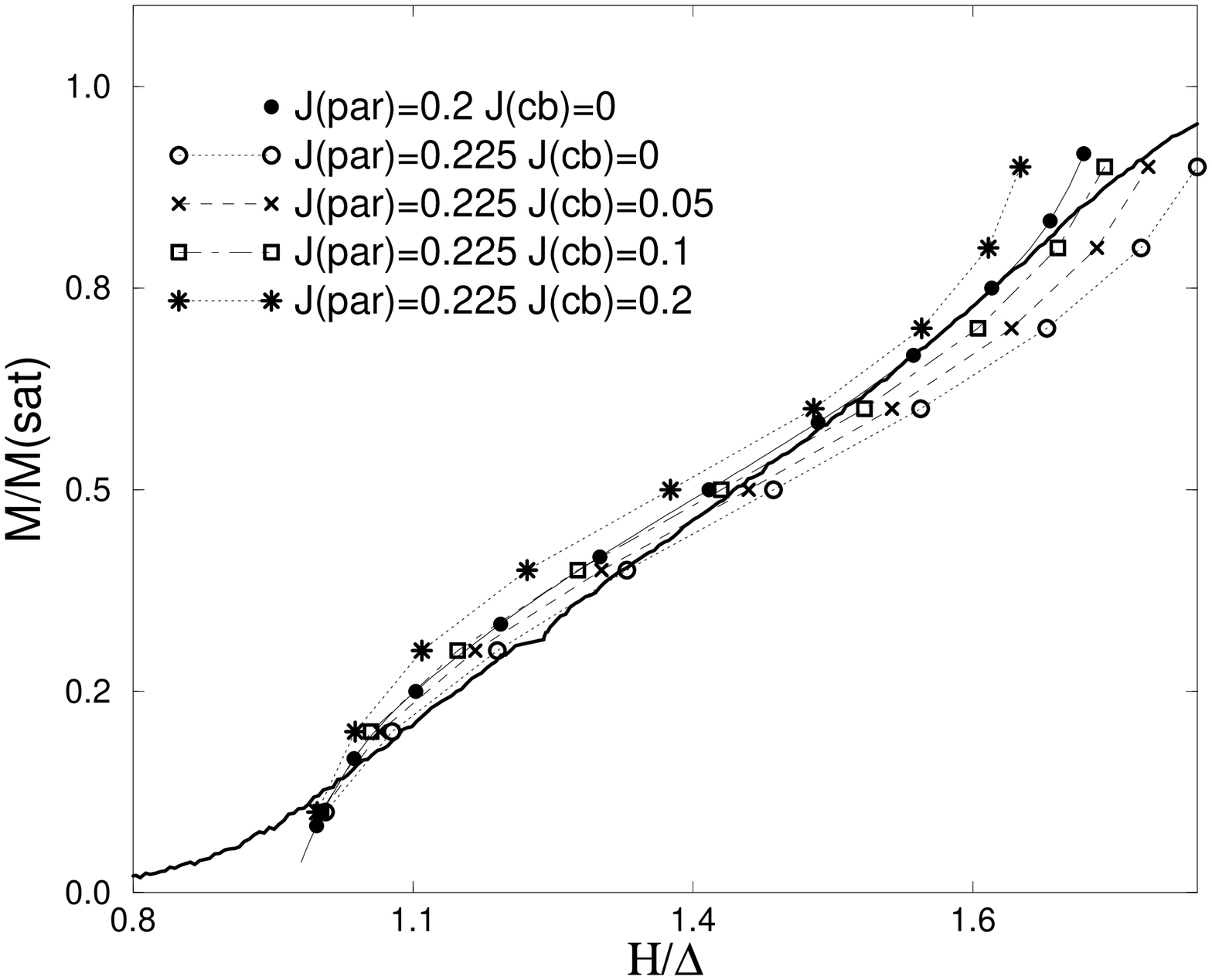,width=\columnwidth,angle=-90}
\end{figure}
Summarising the results above, we find that the Heisenberg model on a 
ladder geometry describes well the magnetization of 
$Cu_2(C_5H_{12}N_2)_2Cl_4$ with a ratio $J/J^\prime\sim 0.2$.
Temperature causes a rounding of the magnetization curves in the vicinity 
of the critical fields and random spin impurities tend to increase
$H_{c2}$. Diagonal bonds improve the theoretical/experimental agreement
further and we suggest that the diagonal interactions appear to be 
ferromagnetic. The discrepancy that still exists however is perhaps not 
surprising since we are using only a simple Heisenberg model. 
Some complications may arise 
due to the fact that in the experimental data provided, the field 
$H_\alpha$ is not applied perpendicular to the plane of the chains
and hence the magnetization ${\cal M}$ is not parallel to the applied field:
there is a $g-$factor anisotropy. 

Another point we should make is that we find  various choices of 
parameters which give reasonably good agreement with the experimental results. 
Fitting the magnetization curve alone is not sufficient to allow 
the relative strengths of the parameters to be fixed extremely 
accurately  
(the Heisenberg chain with both 
nearest and next-nearest neighbour interactions \cite{j1j2} also gives
similar results). 
We can however make reasonable deductions about the effective
interactions present in the compound.

\bigskip
{\it Laboratoire de Physique Quantique, Toulouse} is 
{\it Unit\'e de Recherche Associ\'e au CNRS No 505}. 
CAH and 
DP  acknowledge support from the EEC Human Capital and Mobility 
program under Grants ERBCHBICT941392 and  CHRX-CT93-0332;  
We also thank IDRIS (Orsay) 
for allocation of CPU time on the C94 and C98 CRAY supercomputers.


\begin{references}
\bibitem{note1} If the axial symmetry is broken (eg due to
anisotropy), the  $XY$ correlations
have true long range order.
\end{references}
\end{document}